\documentclass[fleqn,10pt]{wlscirep}
\usepackage[utf8]{inputenc}
\usepackage[T1]{fontenc}
\usepackage{mathtools}
\usepackage{times}
\usepackage{lineno}
\usepackage{xcolor}
\usepackage{xspace}
\usepackage{hyperref}
\usepackage{comment}
\usepackage{xcite, float}
\usepackage{xr}
\usepackage{svg}
\usepackage{xcolor,colortbl}

\title{COVID-19 is linked to changes in the time-space dimension of human mobility}

\author[1]{Clodomir Santana}
\author[1,2]{Federico Botta}
\author[1]{Hugo Barbosa}
\author[3]{Filippo Privitera}
\author[1,2,4]{Ronaldo Menezes}
\author[1,2,5*]{Riccardo {Di Clemente}}

\affil[1]{University of Exeter, Computer Science Department, Exeter, EX4 4QF, United Kingdom}
\affil[2]{The Alan Turing Institute, London, NW1 2DB, United Kingdom}
\affil[3]{Spectus, New York, New York, 10036, United States}
\affil[4]{Federal University of Cear\'a, Fortaleza, 60020-181, Brazil}
\affil[5]{Complex Connections Lab, Network Science Institute, Northeastern University London, London, E1W 1LP, United Kingdom}

\affil[*]{corresponding author: Riccardo.DiClemente@nulondon.ac.uk}

\date{}

\begin{abstract}
Socio-economic constructs and urban topology are crucial drivers of human mobility patterns. During the COVID-19 pandemic, these patterns were re-shaped in their components: the spatial dimension represented by the daily travelled distance, and the temporal dimension expressed as the synchronisation time of commuting routines. Leveraging location-based data from de-identified mobile phone users, we observed that during lockdowns restrictions, the decrease of spatial mobility is interwoven with the emergence of asynchronous mobility dynamics. The lifting of restriction in urban mobility allowed a faster recovery of the spatial dimension compared to the temporal one. Moreover, the recovery in mobility was different depending on urbanisation levels and economic stratification. In rural and low-income areas, the spatial mobility dimension suffered a more significant disruption when compared to urbanised and high-income areas. In contrast, the temporal dimension was more affected in urbanised and high-income areas than in rural and low-income areas.
\end{abstract}
\begin{document} 

\flushbottom
\maketitle

\thispagestyle{empty}

\section*{Introduction}

The places we visit \cite{bohannon2006tracking, andrade2020discovering, Botta2015}, the products we purchase \cite{Xu2019,de2015unique, Bannister2021},  the people we interact with \cite{Candia2008,gao2013discovering,eagle2009inferring}, among other activities, produce digital records of our daily activities. Once decoded and analysed, this digital fingerprint provides a new ground to portray urban dynamics\cite{pentland2013data,lazer2009life}. In particular, the sequence of locations gathered from mobile phone devices via  Calls Detail Records (CDR) and Location-Based Service data (LBS) offers a unique opportunity to assess a broad time span of people's urban activities in almost real-time, overcoming the limitations of surveys and censuses \cite{toole2015,jiang2016timegeo}. CDR and LBS give us information about people's daily motifs across urban locations \cite{Schneider2013}, their attitude in exploring different places \cite{Pappalardo2015}, the route of their commutes \cite{Xu2021,Kalila2018} and the purpose of their urban journey \cite{jiang2017activity}. The location of people in cities is predictable \cite{song2010limits} and is strictly connected with the circadian rhythms of social activities \cite{aubourg2020novel}, as well as home and work locations. The  spatio-temporal variability of commuting patterns \cite{Kung2014} is intertwined with the mode of journey \cite{wang2017human}, the population density (i.e. urbanisation level) \cite{yuan2012extracting}  and the socio-economic status \cite{di2017sequence,pappalardo2015using,barbosa2021uncovering}.

CDR and LBS contribute to the continuous creation of snapshots of citizens' mobility patterns and represent a needed instrument to provide valuable insights on population dynamics in circumstances that urge rapid response \cite{oliver2020mobile,Ubaldi2021}. They have been used to inform public health policymakers assessing the spread of a disease across the population \cite{bengtsson2015using, peak2018population, wesolowski2012quantifying}. Recently, during the COVID-19 pandemic, LBS metrics have become a proxy to evaluate the effectiveness and effects of mobility restriction policies enforced by local governments worldwide \cite{chinazzi2020, Bonaccorsi2020, fraiberger2020uncovering, xiong2020mobile}. Using aggregated mobility data, researchers around the world can develop models to study and predict transmission dynamics \cite{zhou2020effects, da2021meteorological, di2020impact}, investigate the impact and effectiveness of restriction policies, and re-opening strategies \cite{black2020, jones2021impact, wellenius2020impacts, drake2020effects, dahlberg2020effects, yabe2020non, di2020impact,  pullano2020evaluating, schlosser2020covid, pepe2020covid},  and analyse the effects of these policies on the local economy, ethnic and socio-economic groups \cite{showalter2021tribal, gozzi2020estimating, bonato2020mobile, galeazzi2021human, bonaccorsi2021socioeconomic}. Moreover, coupling the mobility data with the socio-economic and ethnicity groups from the census, it is possible to estimate the socio-economic impact of such restrictions in each different community \cite{tizzoni2014use, hou2021intracounty, chang2021mobility, weill2020social,Vaitla2017}.

The majority of the current literature is focusing on the changes in the spatial dimension of mobility  during the COVID-19 pandemic i.e. if citizens are changing the patterns of their whereabouts in terms of magnitude (radius of gyration \cite{iio2021covid} and/or the location visited \cite{pepe2020covid}). In both cases, using de-identified data from mobile phone users, the authors employ the radius of gyration to assess the spatial differences in mobility. The temporal analyses are restricted to assessing trip duration of changes in spatial mobility over time. These studies do not address synchronised mobility patterns or other temporal aspects of human mobility during the pandemic. In addition, both works were published in the early stages of the pandemic, so mobility changes in the same population during different lockdowns could not be studied. Besides the spatial patterns, human whereabouts follow temporal regularities driven by physiology, natural cycles, and social constructs \cite{aledavood2022quantifying}. Few studies have explored these regularities aiming to characterise temporal components and classify people according to weights on these components \cite{aledavood2022quantifying} or to uncover the emergency social phenomena such as the \textit{Familiar Strange} \cite{leng2021understanding}. In the context of the pandemic, it was found that morning activity started later, evening activity started earlier, and temporal behavioural patterns on weekdays became more similar to weekends \cite{sparks2022shifting}. Since urban mobility patterns are built upon the space-time interaction \cite{gao2015spatio}, it is vital also to study both dimensions of mobility to shed light on the mechanisms behind the changes in human mobility during the COVID-19 pandemic.

We can assess the space-time interaction \cite{cmobility} of human activities, studying the rhythms of human mobility with the spatial span of the urban whereabouts. The challenge at hand is to disentangle and investigate how each dimension has been reshaped during the pandemic. To assess the changes in the spatial mobility patterns, preserving citizen privacy under the GDPR (General Data Protection Regulation, more information available at \href{https://gdpr-info.eu/}{gdpr-info.eu/}, accessed on 1 June 2023), we employ the radius of gyration \cite{gonzalez2008understanding, song2010modelling} as a spatial metric. The radius of gyration was chosen for being a well-known metric applied to measure human mobility \cite{du2022lbs, wang2014quantifying, Pappalardo2015, gonzalez2008understanding, song2010modelling, song2010limits, lu2013approaching, wang2017aggregated}. This measure was used during COVID-19 to gauge the general population's compliance with mobility restrictions \cite{haddawy2021effects, di2020impact, pepe2020covid},  inform policymakers on their decisions \cite{pepe2020covid, yabe2020non, di2020impact}, and reveal differences in the impact on different socioeconomic groups and minorities \cite{fan2021fine, gauvin2021socio, reisch2021behavioral, morasae2022economic}.  Besides the regularities in the spatial dimension, human mobility patterns also exhibit a high degree of temporal regularity \cite{gonzalez2008understanding}. These regularities are related to circadian rhythms \cite{aledavood2022quantifying} and commuting for work \cite{zhao2016urban}, study\cite{zhao2016urban} or shopping purposes \cite{oliveira2016regularity}, for example.  In our work, to gauge the temporal dimension of human mobility, we defined the mobility synchronisation metric to quantify the co-temporal occurrence of the daily mobility motifs. It mainly measures the regularities linked to synchronise work schedules, i.e. people leaving home around the same time to go to work. Increased synchronised mobility leads to augmented social contact rates which elevate the risk of transmission of infectious diseases  \cite{leng2021understanding}. Hence, mobility synchronisation can provide relevant insights to policy markers during the pandemic. Combining these spatio-temporal metrics gives us an idea of how far infectious individuals could potentially travel and how many people they could be in contact with (e.g. public transport and office spaces).

Leveraging LBS data from de-identified mobile phone users who opted-in to anonymous location sharing for research purposes, we study how citizen mobility patterns changed from January 2019 to February 2021 across the UK. As the pandemic unfolded, we observed changes in the duration and frequency of trips and disentangled how each mobility dimension was affected. At the lifting of each restrictions, the spatial mobility dimension recovered faster than the temporal dimension. The space-time components drift their trends during the second lockdown to finally align back after the third lockdown. Trips are defined in our paper as the event in which a user leaves their geofenced home area. For each trip, we registered the total time spent outside before returning home, if the trip included a green area, and the distance travelled.

We coupled the mobility dimensions with the urbanisation, unemployment, occupation, and income levels from the census at the local authorities level.  Rural and urban areas manifest opposite trends. In rural areas, the lockdowns affect more the spatial dimension, where the locations of human activities are spread apart and consequently the trips are longer (data from National Travel Survey: England 2018, available at \href{https://www.gov.uk/government/statistics/national-travel-survey-2018}{gov.uk/government/statistics/national-travel-survey-2018}, accessed on 1 June 2023). Meanwhile, in urbanised areas, the synchronicity of the daily activity was dissolved possibly by the rise of asynchronous communing patterns (e.g. flexible work hours or rotational/staggered shifts)  \cite{chung2021covid, watson2020coronavirus}.

We observed that, during the pandemic, the unemployment rates affect more the temporal dimension than the spatial one, where high unemployment levels were associated with low mobility synchronisation. Moreover, this effect seems to be tied with the urbanisation level of the local authorities. Lastly, we adopt the national statistics socio-economic classification (NS-SEC) as a proxy to gauge the impact of the pandemic on the mobility patterns of income/occupation groups. We noticed that areas with elevated concentration of population on low-income routine occupations had the most significant reduction in the spatial and temporal dimensions of mobility.

\section*{Results}

Throughout this section, we define the spatial dimension of human mobility as the span of the citizen's movement, i.e. the length of the trips. This dimension is gauged using the radius of gyration, which quantifies \emph{how far} from the centre of a user's mobility the visited geographical locations are spread. In the temporal dimension, we are interested in measuring co-temporal events linked to collective, synchronised behaviours. This dimension is estimated with the mobility synchronisation metric that represents temporal regularities related to \emph{when} people tend to leave their residences at regular time period. Our goal is to quantify how containment measures  (e.g., limited social gatherings, business and schools closures, home working) affected travel rhythms of the populations. The intervals are identified through the analysis of the strongest frequency components in Fourier spectra of the \emph{out-of-home trips} time series. For this analysis, we use the trip data aggregated hourly. More information on these metrics is available in the Methods section.

We analysed LBS data from January 2019 to February 2021 in the UK. This period includes the three lockdowns announced by the UK's prime minister (GOV.UK: Coronavirus press conferences, available at \href{https://www.gov.uk/government/collections/slides-and-datasets-to-accompany-coronavirus-press-conferences}{gov.uk/government/collections/slides-and-datasets-to-accompany-coronavirus-press-conferences}, accessed on 1 June 2023). Given the discrepancies in the implementation of the different lockdown across each country in the UK, an analysis of the local/regional impact of the pandemic is included in the Supplementary Information: Northern Ireland(Supplementary Figure \textbf{7}), Scotland (Supplementary Figure \textbf{8}), Wales (Supplementary Figure \textbf{9}) and England (Supplementary Figure \textbf{6}).

Fig. \textbf{1} depicts the radius of gyration and mobility synchronisation trends from the second week of 2020 to seventh week of 2021. Both metrics have similar trends up to week 18 of 2020 when the radius starts recovering to pre-pandemic levels while the synchronisation does not. The recovery in the spatial dimension coupled with the fluctuations in the temporal patterns suggests that, although people gradually started making trips similar to the period before the pandemic, these trips do not present the temporal synchronisation observed before. After the third lockdown, we can notice similar trends in the spatial and temporal dimensions as in the period before week 18.

\begin{figure}[!ht]
\centering
\includegraphics[width=0.95\textwidth]{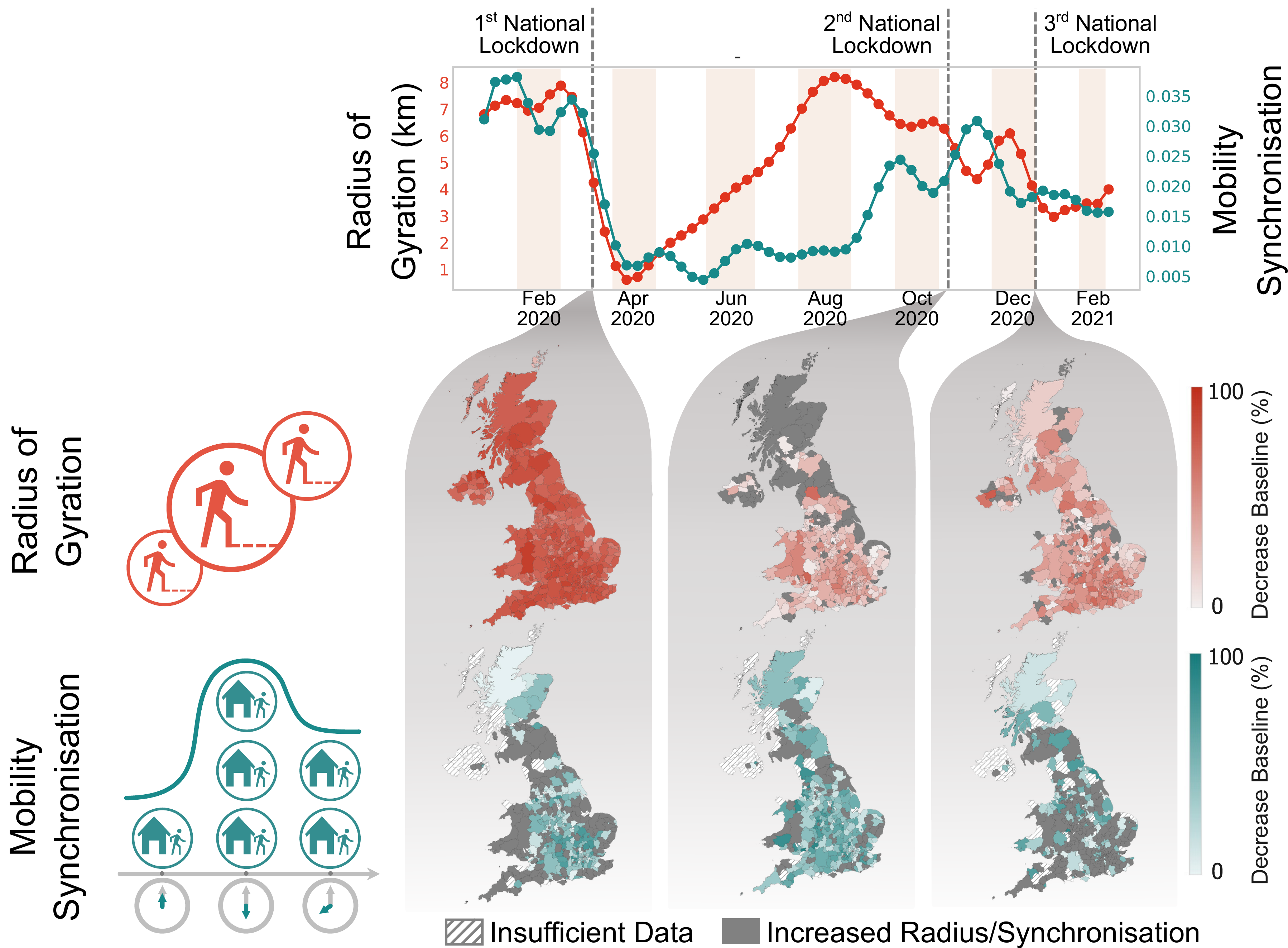}
\caption{Evolution of the radius of gyration and mobility synchronisation in the UK’s local authorities from the second week of 2020 to week 7 of 2021. The maps depict the effects of the three English national lockdowns regarding the spatio-temporal metrics adopted. Note that the first lockdown resulted in the most significant reduction in the radius of gyration in all four nations. However, it is worth mentioning that the first lockdown was the only one with the same restrictions for all UK’s countries. Concerning mobility synchronisation, the most notable reduction occurred during the second lockdown. Additionally, it is challenging to disentangle the effects of the third lockdown from the changes in the population mobility patterns caused by end-of-year holidays.}
\label{fig:figure_1}
\end{figure}

Since mobility synchronisation measures the existing time trends, it can estimate the effects of human mobility restrictions policies, such as lockdowns. We can see reductions in the mobility synchronisation levels during all three lockdowns. However, the second one produced the most significant decrease in the synchronicity levels in most local authorities compared to the other two; 78.34\% of the local authorities experienced a reduction in the synchronisation level compared to the baseline. In contrast, during the first and the third lockdown,  the mobility synchronisation decreased in $56.67\%$ and $34.02\%$ of the local authorities, respectively. The substantial drop in the mobility synchronisation during the second lockdown might be tied to the changes in the \emph{stay-at-home} policy \cite{england2020health, world2020public, world2020global}. This policy mainly only allowed essential workers to leave home to work, while in the second and third lockdowns this rule became more flexible and people who could not work from home were allowed to go to work \cite{england2020health, scotland2020health, ireland2020health, wales2020health, scally2020uk}.

\subsection*{Changes in mobility according to urbanisation level and economic stratification}

Human mobility patterns are also affected by urbanisation level \cite{alessandretti2020scales}. For example, rural areas are characterised by limited accessibility to goods, services and activities \cite{berry1967geography}. In the context of a pandemic, the urbanisation level partly explains differences in mobility patterns \cite{kishore2021lockdowns}, and the diffusion of infectious diseases \cite{sigler2021socio, neiderud2015urbanization, ali2006global}.

To assess how the spatio-temporal mobility patterns of areas with different levels of urbanisation were affected during the COVID-19 pandemic, we analyse the radius of gyration and mobility synchronisation by urbanisation level. We divided the local authorities into three classes accordingly to the urban-rural classification adopted for England \cite{pateman2011rural} illustrated in Fig. \textbf{2a}.

\begin{figure}[!ht]
\centering
\includegraphics[width=0.95\textwidth]{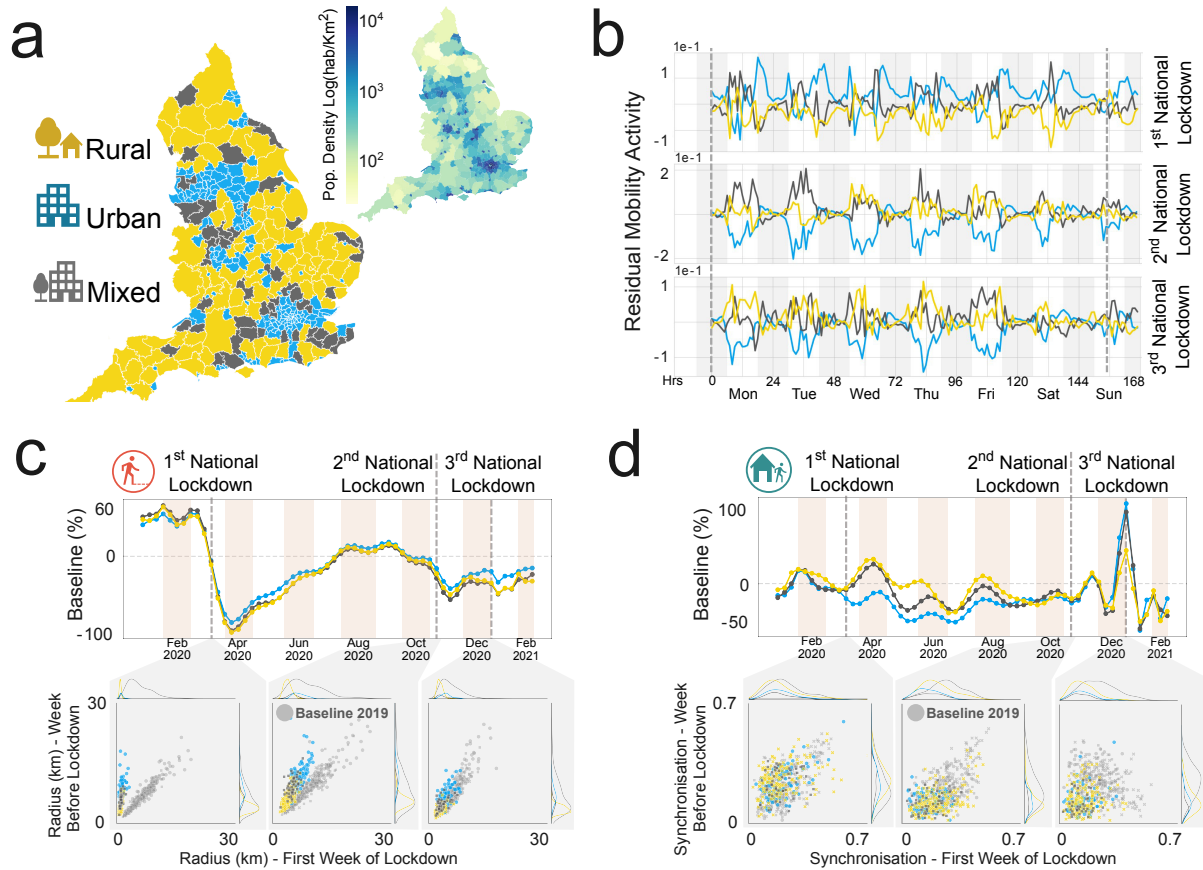}
\caption{Radius of gyration and mobility synchronisation of English local authorities grouped according to the Urban-Rural classification (a). (b) compares the differences in the number of out-of-home trips for the different urban-rural groups. Notice that after the first lockdown, urban local authorities started to present negative values in their curve. (c) and (d) illustrate, respectively, a time series with variations on the radius of gyration and mobility synchronisation when compared to the baseline year (2019).}
\label{fig:figure_2}
\end{figure}

We employed the concept of residual activity \cite{gonzalez2008understanding} to visualise the deviations of the mobility patterns during the lockdowns when compared to their expected behaviour (e.g. we used the same period of 2019 as the baseline for comparisons). Fig. \textbf{2b} depicts the different responses of the urban-rural group to the three national lockdowns. High residual values indicate an increased number of trips compared to the expected behaviour (baseline period).

During the first lockdown, urban areas presented an increase in expected mobility. In contrast, rural areas have a negative trend. However, during the second and third lockdown, an opposite scenario emerges. Rural local authorities increased the expected residual activity, and urban areas decreased it.  These differences can be driven by the change in the mobility restriction policies (e.g. more flexible stay-at-home rules \cite{england2020health}) and the characteristics and pre-existing social vulnerabilities of urban and rural areas, as found in previous works \cite{huang2021urban}

Although there is a debate as to whether a high population density accelerates or not the spread of the virus \cite{khavarian2021high}, other urban and rural characteristics can be risk factors for COVID-19. For example, transportation systems and increased inter/intra urban connectivity are regarded as key factors contributing to the spread of contagious diseases \cite{kutela2021exploring}.

The results of the radius of gyration (Fig. \textbf{2c}) and the mobility synchronisation (Fig. \textbf{2d}) also indicate differences in the response of urban and rural areas to the lockdowns. Due to the characteristics of the geographic distribution of local amenities in rural areas, people tend to have a greater radius of gyration compared to urban areas (data from National Travel Survey: England 2018, available at \href{https://www.gov.uk/government/statistics/national-travel-survey-2018}{gov.uk/government/statistics/national-travel-survey-2018}, accessed on 1 June 2023).

Analysing the trends of the radius of gyration and mobility synchronisation compared to the baseline of 2019, we can notice some differences in the urban-rural and spatio-temporal response of the local authorities (scatter plots on Fig. \textbf{2c} and Fig. \textbf{2d}). In the spatial dimension, we can see the same behaviour for urban and rural areas, characterised by a reduction in the mobility levels in the week before and the first week of lockdown. The shift between the baseline and the first/second lockdown indicates that the radius during the first week was significantly smaller than the week before these lockdowns.

In the temporal dimension, however, the differences between the mobility levels in the week before the lockdown and the first week of lockdown (first scatter plot in Fig. \textbf{2d}) are less dramatic than observed in the spatial dimension. Nonetheless, compared to the baseline, we can still see a reduction in the synchronisation values for urban and rural areas, especially in the second and third lockdowns (baseline plot in Fig. \textbf{2d}).

As discussed, the number of trips has decreased across all the urban-rural groups during the pandemic. Since work-related activities often create the necessity to leave home, the rise in home-working and unemployment rate contribute to the reduction in the mobility levels \cite{lee2020human}. Next, we study the relationship between the spatio-temporal mobility metrics and the unemployment rate for areas with different levels of urbanisation, both before and during the pandemic.  We estimate the size of the unemployed population based on the unemployment claimant count (Data from the Office for National Statistics, available at \href{https://www.ons.gov.uk/employmentandlabourmarket/peoplenotinwork/unemployment/datalist?filter=datasets}{ons.gov.uk/employmentandlabourmarket}, accessed on 1 June 2023).

We divided the time series of the radius of gyration and mobility synchronisation into pre-pandemic (from  April 2019 to February 2020) and pandemic (from  April 2020 to February 2021) periods, and we analysed the Kendall Tau correlation between them and unemployment claimant count.

\begin{figure}[!ht]
\centering
\includegraphics[width=0.95\textwidth]{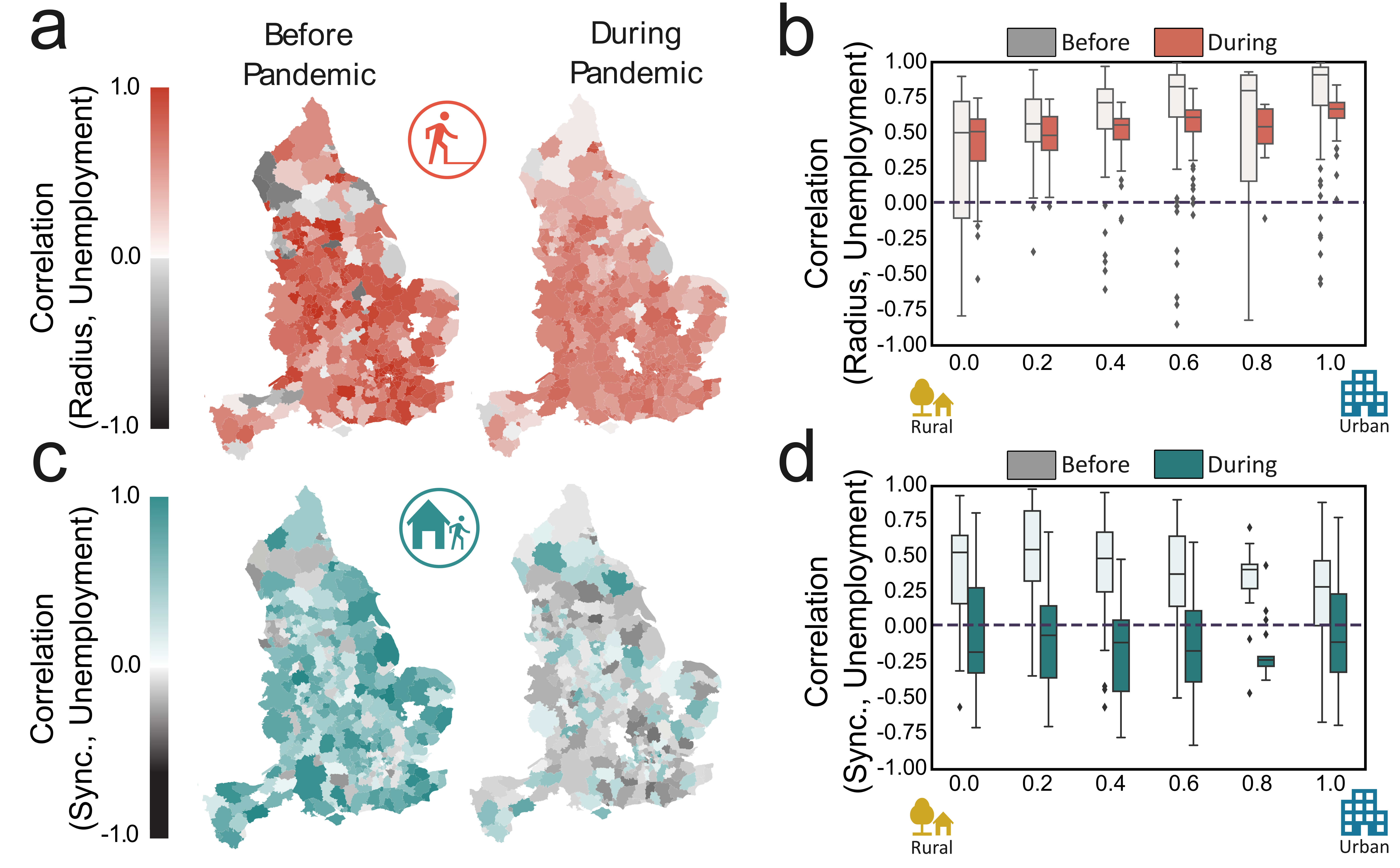}
\caption{Correlation between the spatio-temporal metrics of human mobility and the unemployment claimant count. (a) and (b) look at the correlation between the radius of gyration and mobility synchronisation with the unemployment claimant count for the periods before and during the pandemic ($N=316$ English local authorities). We define the period before the pandemic from April 2019 to February 2020. The pandemic period considered was from April 2020 to February 2021. (b) and (d) consists of boxes that span from the 25th percentile (Q1) to the 75th percentile (Q3), with median values (50th percentile) represented by a central line. The minimum and maximum values are determined by subtracting 1.5 times the interquartile range (IQR) from Q1 and adding 1.5 times the IQR to Q3, respectively. Notice that, before the pandemic, both the radius and the synchronisation positively correlated with the unemployment claimant count. However, during the pandemic, the correlation with the radius became less strong, and that with synchronisation became negative. It is also worth mentioning that (b) and (d) reveal a possible association between correlation changes and areas’ urbanisation level.}
\label{fig:figure_3}
\end{figure}

At first glance, the positive correlation between the radius of gyration and the unemployment rate (Fig. \textbf{3a} and Fig. \textbf{3b}) seems to be dissonant from previous works \cite{Toole2015a, pappalardo2015using}. However, this result can be due to a rise in the unemployment rate and the spatial mobility levels before the pandemic. For the pandemic period, although the lockdowns have reduced the radius during specific periods, we see in Fig. \textbf{1} a period between April and August 2020 when the radius increased, making the correlation positive.

While the correlation between the spatial dimension of mobility and unemployment drops only marginally, preserving the positive sign and affecting more urban than rural areas, the correlation between the time dimension of mobility and  unemployment drops significantly, becoming negative and impacting more rural districts than the urban, see (Fig. \textbf{3c} and Fig. \textbf{3d}).

The same explanation applies to mobility synchronisation in the pre-pandemic period. During the pandemic, we can see that the spatio-temporal dimensions display different patterns between May and November, and the temporal one does not present the same steep recovery as the spatial between April and August 2020. The oscillations in the patterns of the temporal dimension impacted the correlation with unemployment, making it negative.

We argued that work trips contribute to the creation of our mobility patterns. Moreover, the type of occupation, among other socio-demographic characteristics,  also influences those patterns \cite{Lenormand2015}. In this sense, we use the English National Statistics Socio-economic classification (NS-SEC) (The National Statistics Socio-economic Classification, available at \href{https://www.ons.gov.uk/methodology/classificationsandstandards/otherclassifications/thenationalstatisticssocioeconomicclassificationnssecrebasedonsoc2010}{ons.gov.uk}, accessed on 1 June 2023) to gauge the response of employment relations and conditions of occupations to the pandemic. Using this classification, we also have insights into the connection between the spatio-temporal mobility and wealth level.  This analysis is important since wealth and economic segregation are linked to differences in mobility patterns and response to human mobility restrictions under COVID-19 pandemic \cite{Bonaccorsi2020}.

In the NS-SEC classification, classes related to managerial occupations exhibit strong positive correlation with income as depicted in Fig. \textbf{4a}. In contrast, lower supervisory, technical, semi-routine or routine occupations negatively correlate with income. The remainder of NS-SEC classes they present a strong correlation with income.

\begin{figure}[!ht]
\centering
\includegraphics[width=0.95\textwidth]{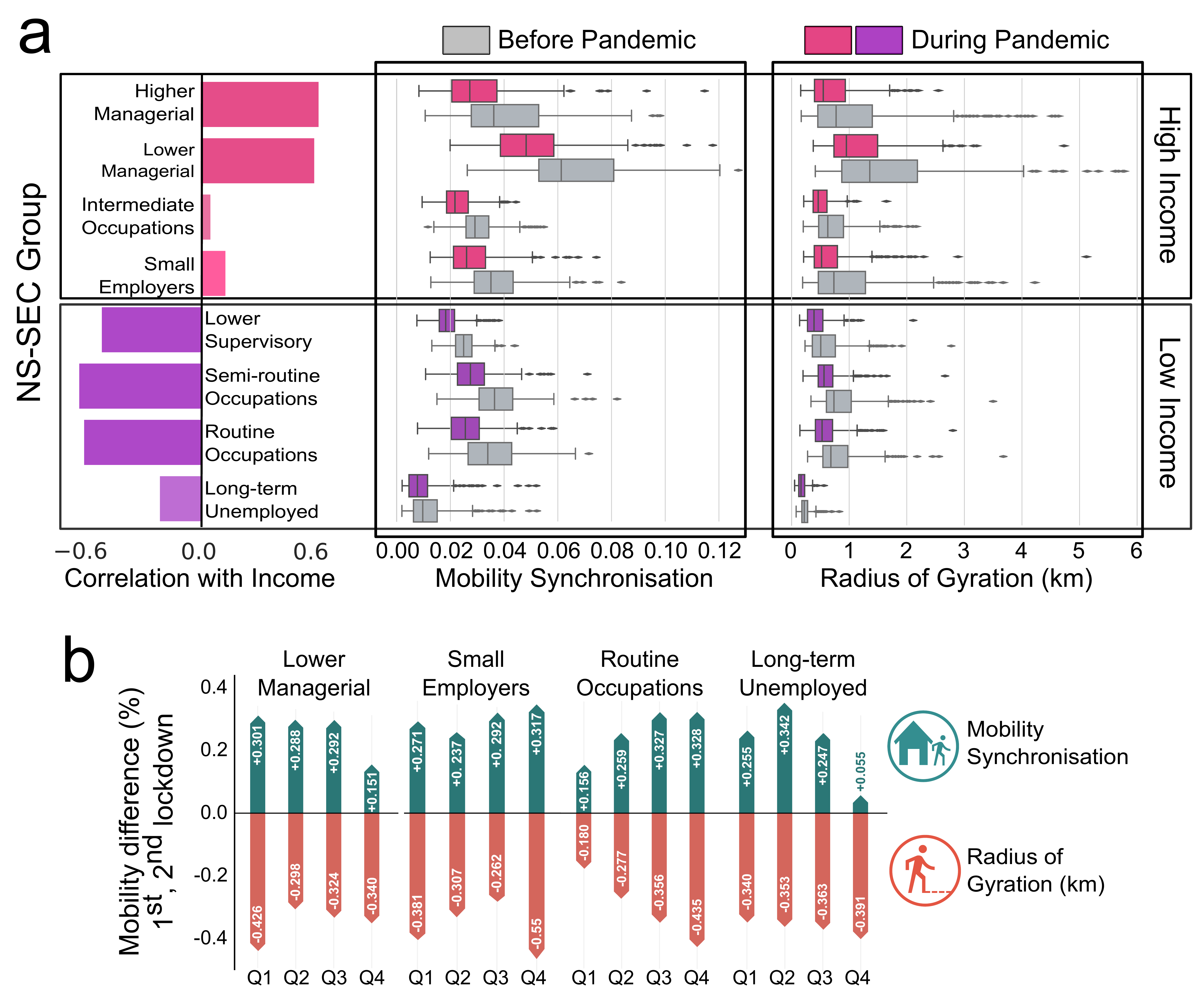}
\caption{Relation between the NS-SEC classification, income and their impact on the spatio-temporal mobility patterns before and during the pandemic. The “Before the Pandemic" label corresponds to the period from April 2019 to February 2020. While the “During Pandemic" one happened between April 2020 and February 2021. (a) looks at the correlation between the NS-SEC classes and the population income ($N= 316$ English local authorities). In this panel, the classes coloured in pink and purple have positive and negative correlations with the population income. The box plot in (a) consists of boxes that span from the 25th percentile (Q1) to the 75th percentile (Q3), with median values (50th percentile) represented by a central line. The minimum and maximum values are determined by subtracting 1.5 times the interquartile range (IQR) from Q1 and adding 1.5 times the IQR to Q3, respectively. (a) also shows the radius of gyration/mobility synchronisation of each NS-SEC class before and during the pandemic. Lastly, (b) highlights the differences between the first and second lockdowns in the spatio-temporal metrics.}
\label{fig:figure_4}
\end{figure}

Moreover, Fig. \textbf{4b} shows that, before and during the COVID-19 pandemic, people's radius of gyration and the mobility synchronisation within an area tend to grow as the concentration of population in managerial occupations increases. The opposite result is observed when analysing the percentage of the population (quartiles Q1 to Q4) in lower supervisory, semi-routine or routine occupations.

In all scenarios depicted in Fig. \textbf{4}, we can see a reduction in the temporal and spatial dimensions of mobility during the pandemic. However, each group contributed differently to the overall change in the mobility observed during the pandemic. As mentioned before, the first national lockdown produced the most significant impact on the spatial dimension of mobility. In contrast, the second greatly impacted the temporal dimension.

Besides the difference in the time-space facets of human mobility analysed, changes in the duration \cite{padmanabhan2021covid} and the purpose \cite{fatmi2020covid} of the trips were also observed during the pandemic. Researchers found that these changes vary accordingly to income levels \cite{fatmi2020covid} and could be related to the emergence of new habits \cite{sheth2020impact}.

To assess the changes in the duration of the trips, we measured the time elapsed when the user left their home geo-fencing area and entered it
again. We can further disaggregate the trips by classifying them as work-related and other types based on their starting time. Comparing the duration of the trips in a week with no mobility restriction Fig. \textbf{5a} to a week with lockdown Fig. \textbf{5b}, we can see that trips classified as work-related display a reduction in their length. The analysis based on the income/socioeconomic groups shows that high-income groups presented the most significant reduction compared to the baseline year of 2019, depicted in Fig. \textbf{5b}. This result is in line with a similar paper in the literature, which also reports differences related to the income groups \cite{sparks2022shifting, padmanabhan2021covid}.

\begin{figure}[!ht]
\centering
\includegraphics[width=0.95\textwidth]{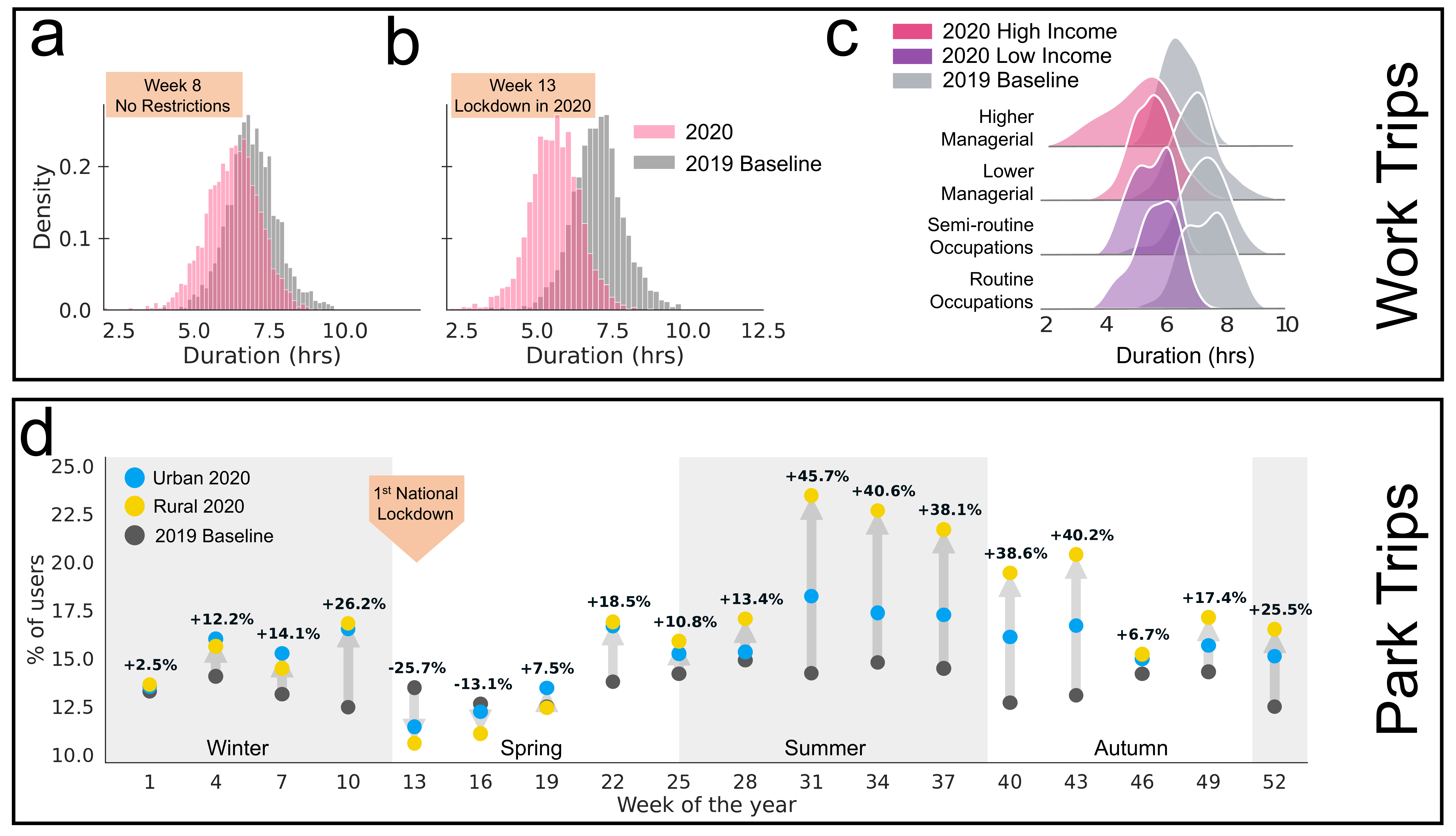}
\caption{Changes in the number and duration of trips. (a) and (b) illustrate the differences between the duration of work-related trips for two weeks in 2019 and 2020. During week 8 (a), there were no mobility restrictions in 2020, and we can see that the distributions are similar. In contrast, week 13 (b) was the first week of the first lockdown, and we can see more differences in the distributions of 2019 and 2020. (c) looks into the differences across different socioeconomic groups during week 13. Lastly, (d) depicts the differences in the number of trips to green spaces such as parks in 2020 compared to the baseline year of 2019. Note that in this case, the local authorities are grouped in urban and rural groups}
\label{fig:figure_5}
\end{figure}

Besides the trips' duration, another relevant aspect being analysed is the impact on the type of trips. We study the differences in leisure-related trips before and during the pandemic to assess this impact. Here, these trips are estimated based on examining trips that include green areas such as parks, sports facilities and play areas. In a period without any mobility restriction measures, we expect to see an increase in these trips from the end of winter until the end of summer. This number should start decreasing as when the winter approaches. As shown in Figure \textbf{5d}, using the baseline year of 2019,  before the first national lockdown, there was no substantial difference between the number of trips by rural and urban local authorities. This pattern stays consistent until the start of the summer when the restrictions are stated to be lifted \cite{england2020health}. After that, we can see a substantial difference in the amount of green areas trips in rural and urban areas. Although one could argue that this could indicate a preference towards more natural leisure trips in rural areas, further investigation is required to identify the reasons behind these differences.

\section*{Discussion}

The accurate estimation of spatio-temporal facets of human mobility and gauging its changes as a reflection of the pandemic and mobility restriction policies is critical for assessing the effectiveness of these policies and mitigating the spread of COVID-19 \cite{england2020health, wales2020health, ireland2020health, scotland2020health, ahmed2020inequality}. A vital contribution of our work lies in applying two metrics to disentangle the changes in mobility's temporal and spatial dimensions. Using the radius of gyration, we could identify that the effect of the first lockdown was more significant than the other ones in changing the spatial characteristics of citizens' movement. Among the reasons that could lead to this result, we can mention more strict policies adopted in the first lockdown and the lockdown duration \cite{scally2020uk, england2020health}. However, further investigation is needed to obtain more pieces of evidence to support these hypotheses.

In contrast to the spatial dimension of mobility, the results indicate that the temporal one was more impacted during the second lockdown when more flexible mobility restriction policies were enforced. After the first lockdown, we argue that people who could not work from home were allowed to leave home and work in the office as long as they respected social distancing rules \cite{scally2020uk, world2020global}. Different work shifts were created to comply with these rules and limit the number of people inside indoor spaces. As a result, instead of having the same, or similar, work schedule for all employees, they started to be divided into groups that should go on different days of the week and at different times, impacting the temporal synchronisation of their movement.

Besides the changes during different periods and under different mobility restriction policies, we also analysed the interplay between the characteristics of the area (e.g. level of urbanisation, income and unemployment rates) and the spatio-temporal human mobility metrics adopted. We noticed that rural areas presented a more significant reduction in spatial patterns compared to urban ones. At the same time, urban areas were more impacted in their temporal dimension than rural ones. These differences observed in the urban-rural classification are correlated to the population density of the regions and can affect the impact of the mobility restriction measures.

We also observed changes in the response of regions regarding their unemployment rates and populations income. For the unemployment rates, we observed that before the COVID-19 pandemic, the unemployment claimant count was positively correlated with the radius of gyration and mobility synchronisation in the majority of the areas. However, during the pandemic, this correlation became weaker for the radius of gyration and negative for mobility synchronisation. Less urbanised areas tend to have a lower spatial correlation and a higher temporal correlation with unemployment than more urbanised areas. Using the NS-SEC classification as a proxy to assess the response of different income and work groups, we observed that low-income routine/semi-routine occupations were the groups that presented the greatest reduction in their radius and synchronisation. Moreover, the changes in the mobility restriction policies after the first lockdown had an impact on these groups, which can be the reason behind the greatest impact on the temporal dimension of mobility \cite{scally2020uk, england2020health}. Similarly, changes were also observed concerning the duration of work-related trips and the number of trips to green areas. These differences were also observed at urbanisation and socioeconomic levels.

Regarding our work's contribution to policy-makers, we argue that the spatio-temporal metrics employed in this study help to assess mobility changes before and after the implementation of policies, as seen in the first and second lockdowns with flexible \emph{stay-at-home} policies \cite{england2020health, world2020public, world2020global}. Moreover, our results indicate that different groups (socioeconomic and urban-rural) experience and respond to these policies differently. These results were also found by similar works \cite{padmanabhan2021covid, lee2020human, huang2021urban ,morasae2022economic} and provide insight to policy-makers to design strategies that consider each group's particularities.

In summary, the analysis of the spatial dimension of human mobility coupled with the insights from the study of the temporal dimension allows us to characterise the impact of policies such as \emph{stay-at-home} and school closures on the population of different areas/socioeconomics. These differences suggest that each group experiences, in a particular way, the emergence of asynchronous mobility patterns primarily due to the enforcement of mobility restriction policies and new habits (e.g. home office and home education).

\section*{Methods}

\subsection*{Data Sources} 

\subsubsection*{Human Mobility Data} 

Spectus Inc provided the human mobility data used in this research for research purposes. This data was collected from anonymous mobile phone users who have opted-in to give access to their location data anonymously, through a GDPR-compliant framework. Researchers queried the mobility data through an auditable, cloud-hosted sandbox environment, receiving aggregate outputs in return. The datasets contain records of UK users from January 2019 to early march 2021. In total, we have over 17.8 billion Out-of-home trips and about 1 billion users' radius of gyration records. Note that the radius logs are measured on a weekly basis while the trips are recorded on a daily basis. More information on the datasets is available in the SI. We assessed the representativeness of the data by analysing the correlation between the number of users and the population of the local authorities. A strong positive correlation between the populations compared with $r^2$ value equal to 0.775 was obtained (see the Supplementary Figure \textbf{1a}). The data is composed of records of when users leave the area (as a square of 500m) that is related to their homes and the length of the trips. The home area is estimated based on the history of places visited, the amount of time spent at the area and the period when the user stayed at the location following\cite{Hariharan2004,Kung2014}. It is worth mentioning that we only used mobility-related data in this work due to privacy concerns. No personal information or another type of information that would allow the identification of uses was used. Also, the data is aggregated at the local authority level as the aim is to study trends at the population level rather than the individuals. The dataset on the trips to the green areas is composed of records of the number of trips which include green spaces across Great Britain. These records were collected daily grouped by the hour the trip starts, and are aggregated at the local authority level.

\subsubsection*{Other Data Sources}

The Income study employed data from the report on income estimates for small areas in England in 2018 provided by the Office for National Statistics (ONS) \cite{eng2008income}. It is available for download and distribution under the terms of the Open Government Licence (available at \href{http://www.nationalarchives.gov.uk/doc/open-government-licence/version/2}{www.nationalarchives.gov.uk/}, accessed on 1 June 2023). Similarly, the analysis of the unemployment claimant rate, the urban-rural classification of English local authorities \cite{pateman2011rural} and the national statistics socio-economic classification \cite{chandola2000new} used the data published by the ONS publicly available under the terms of the Open Government Licence. The remaining socioeconomic data utilised aggregated data from the UK Census of 2011 \cite{uk2017census}, which is available for download at the InFuse platform also under the terms of the Open Government Licence. For the green area study, we used the Ordnance Survey Open Greenspace dataset to obtain information on the locations of public parks, playing fields, sports facilities and play areas(OS Open Greenspace, available at \href{https://beta.ordnancesurvey.co.uk/products/os-open-greenspace\#technical}{ordnancesurvey.co.uk/os-open-greenspace}, accessed on 1 June 2023). Our analysis did not include categories related to religious grounds, such as burial grounds or churchyards.

\subsection*{Metrics and Other Methods}

\subsubsection*{Radius of Gyration}

We conducted the study of the radius of gyration ($RG$) following the definition of Gonzalez et al. \cite{gonzalez2008understanding}. It can be described as the characteristic distance travelled by a user $u$ during a period and is calculated as follows

\begin{equation}
    RG_u = \sqrt{ \frac{1}{N_u} \sum_{i=1}^{N_u} (\vec{r}_{\,u}^{\,i} - \vec{r}_{\,u}^{\,cm})^2}
\end{equation}

\noindent where, $N_u$ represents the unique locations visited by the user, $\vec{r}_{\,u}^{\,i}$ is the geographic coordinate of location $i$ and $\vec{r}_{\,u}^{\,cm}$ indicates the centre of mass of the trajectory calculated by

\begin{equation}
    \vec{r}_{\,u}^{\,cm} = \frac{1}{N_u} \sum_{i=1}^{N_u} n_{\,u}^{\,i}\vec{r}_{\,u}^{\,i}
\end{equation}

\noindent where $n_{\,u}^{\,i}$ is the visit frequency or the waiting time in location $i$. The mobility value of each region is the median value of the radius of gyration of the users within a temporal window of 8 days centred around a given day.

\subsubsection*{Residual Mobility Activity}

The concept of residual mobility activity displayed in Fig. 2 B was adapted from \cite{toole2012inferring}, and it is used to highlight differences between the measured behaviour of the local authorities compared to their expected behaviour. For a given local authority $i$ is calculated as follows

\begin{equation}
    a_{i}^{res}(t) = a_{i}^{norm}(t) - a^{-norm}(t)
\end{equation}

\noindent where $a^{-norm}(t)$ is the normalised activity averaged over all local authorities under at each particular time, and $a_{i}^{norm}(t)$ is computed similarly to the Z-score metric

\begin{equation}
    a_{i}^{norm}(t) = \frac{a_{i}^{abs}(t) - \mu^{abs}}{\sigma^{abs}}
\end{equation}

\noindent where $a_{i}^{abs}(t)$ is the activity in a local authority at a specific time $t$, $\mu^{abs}$ is the mean activity of all local authorities under the same urban-rural classification of $i$ at a specific time, and $\sigma^{abs}$ represents the standard deviation of all local authorities under the same urban-rural classification of $a_{i,j}$ at a particular time. 

\subsubsection*{Mobility Synchronisation}

The mobility synchronisation is not limited to conventional commuting routines. It can happen at any time of the day in which people tend to perform certain activities. For example, school teachers, healthcare professionals,  and other routine or semi-routine occupations tend to have defined times reserved for specific activities (e.g., eating, exercising, socialising). To have a more accurate portrait of the mobility synchronisation patterns, instead of analysing it as a concentration of trips around certain hours, we define the mobility synchronisation as the total magnitude in the periodicity in the out-of-home trips.

First, using 2019 as a baseline, we analyse the Wavelet and Fourier spectra to determine the expected strongest frequency components in the mobility regularity. For a given mother wavelet $\psi(t)$, the discrete wavelet transform can be described as

\begin{equation}
    \psi_{j, k}(t) = \frac{1}{\sqrt{2^j}} \psi \left ( \frac{t - k2^j}{2^j} \right )
\label{eq:wavelet}
\end{equation}

\noindent where $j$ and $k$ are integers that represent, respectively, the scale and the shift parameters. For the Fourier transform, a discrete transform of the signal $x_n$, for $n = 0 \dots N-1$ is:  

\begin{equation}
    X_{k} = \sum_{n=0}^{N-1} x_{n}e^{-i2\pi kn/N}
\label{eq:fourier}
\end{equation}

\noindent where $K = 0 \dots N-1$ and $e^{-i2\pi kn/N}$ represents the \emph{Nth roots of unity}. 

Employing these two transforms, we found that the mobility patterns are characterised by five main periods, namely 24h, 12h, 8h, and 6h (see SI material Supplementary Figure \textbf{3}). However, because the 24h component overshadows the other three components (see Supplementary Figure \textbf{5}), we focus the analysis on the 12h, 8h, and 6h periods. Moreover, during the pandemic, these periods were more affected than the 24hrs component (see Supplementary Figure \textbf{4}). Next, the mobility synchronisation metric is defined as the sum of the powers from the Lomb-Scargle periodograms \cite{vanderplas2018understanding} corresponding to the 12h, 8h and 6h. The generalised Lomb-Scargle periodograms is calculated as
 
 \begin{equation}
     P_{N}(f) = \frac{1}{\sum_{i}y_{i}^{2}}\left \{ \frac{\left [ \sum_{i} y_{i} cos\omega(t_{i} - \tau)   \right ]^{2}}{ \sum_{i} cos^{2}\omega(t_{i} - \tau) } + \frac{\left [ \sum_{i} y_{i} sin\omega(t_{i} - \tau)   \right ]^{2}}{ \sum_{i} sin^{2}\omega(t_{i} - \tau) } \right \}
 \end{equation}
 
 \noindent where $y_i$ represents the N-measurements of a time series at time $t_i$, $\omega$ is a frequency, $\tau$ can be obtained from
 
  \begin{equation}
     tan2\omega\tau = \frac{\sum_{i} sin 2 \omega t_{i}}{\sum_{i} cos 2 \omega t_{i}}
 \end{equation}
 
 The mobility synchronisation is a value between 0 and 1, where higher values for a given period indicate that more people left their homes simultaneously. In the context of the pandemic, high mobility synchronisation can be translated into a potential increase in the likelihood of being exposed or exposing more people to the virus due to the large number of people moving simultaneously.

\section*{Peer-reviewed  Version}

The peer-reviewed version of this pre-print is available at \href{https://www.nature.com/articles/s41562-023-01660-3}{https://www.nature.com/articles/s41562-023-01660-3}. If you would like to reference this work, please cite our paper: Santana, C., Botta, F., Barbosa, H. et al. COVID-19 is linked to changes in the time–space dimension of human mobility. Nat Hum Behav (2023). https://doi.org/10.1038/s41562-023-01660-3

\section*{Data Availability}

The paper contains all the necessary information to assess its conclusions, including details found in both the paper and the Supplementary Materials. Due to contractual and privacy obligations, we are unable to share the raw mobile phone data. However, access can be provided by Spectus Inc upon agreement and signature of the NDA. More information on data access for research can be found at \href{https://spectus.ai/social-impact/#:~:text=Powering%20research%20%26%20academic%20teams%20globally,use%20of%20location%2Dbased%20data.}{Spectus - ``Data for Good" movement}.

\section*{Code Availability}

Scripts and Notebooks in Python with our analyses and to reproduce the results in this paper were archived with Zenodo (\href{https://doi.org/10.5281/zenodo.8014785}{https://doi.org/10.5281/zenodo.8014785}).

\section*{Acknowledgments}

This work is a collaboration between the Department of Computer Science of the University of Exeter and Spectus response to the COVID-19 crisis, \href{https://spectus.ai/}{Spectus} is providing insights to academic and humanitarian groups through a multi-stakeholder \href{https://www.cuebiq.com/about/data-for-good}{data collaborative} for timely and ethical analysis of aggregate human mobility patterns. This manuscript is the outcome of this collaboration Spectus and summarises the main findings. Besides this paper, two additional, non-peer reviewed, reports were produced during the first national lockdown with an initial analysis of the pandemic in the UK and a second one with analysing changes in socio-economic aspects related to mobility patterns in the UK during the first lockdown. Both reports are available at the project web page \href{https://covid19-uk-mobility.github.io/}{covid19-uk-mobility.github.io}. RDC. acknowledges Sony CSL Laboratories in Paris for hosting him during part of the research. FB was funded by the Economic and Social Research Council (ESRC) \& ADR UK as part of the ESRC-ADR UK No.10 Data Science (10DS) fellowship (grant number ES/W003937/1). The funders had no role in study design, data collection and analysis, decision to publish or preparation of the manuscript

\section*{Author Contributions Statement}

CS performed the analysis. RDC and FP gather the data. RDC, FB, HB, RM designed the analysis. CS and RDC wrote the paper. RM, RDC supervised the project. All authors discussed the results and contributed to the final manuscript.

\section*{Competing Interests Statement}

The authors declare no competing interests.


\cleardoublepage
\renewcommand{\thesection}{S\arabic{section}}
\renewcommand{\thefigure}{S\arabic{section}.\arabic{figure}}
\renewcommand{\thetable}{S\arabic{section}.\arabic{table}}
\renewcommand{\theequation}{S\arabic{section}.\arabic{equation}}

\section*{\Huge{\textbf{Supplementary Materials}}}
\setcounter{figure}{0}
\setcounter{table}{0}
\setcounter{equation}{0}

\section{Data Information}
\label{sec:data}

The content of the datasets used in this work that contains information about the radius of gyration, out-of-home trips, trips to green areas, and duration of the trips is described as follows.

\subsection{Radius of Gyration}

Description of the data used to analyse the spatial dimension of mobility, i.e. radius of gyration. Each item corresponds to a column in the dataset.

\begin{description}
    \item[year:] Year when the location information was saved in the database.
    \item[week:] Week number according to the ISO-8601 standard where weeks start on Monday. It is a number between 1 and 52 (or 53 for leap years).
    \item[radius:] The average radius of gyration in kilometres for the users inside a region at a given week of the year. The values were normalised using a min-max normalisation to comply with data sharing policies.
    \item[pings:] Represents the number of pigs (sync connections sent from the users' devices to the data server). The values were also normalised using a min-max normalisation to comply with data sharing policies.
    \item[geo\_code:] Unique codes for local authority districts (LAD) and unitary authorities (UA) in the United Kingdom.
\end{description}

\subsection{Out-of-home Trips}

Description of the data used to analyse the temporal dimension of mobility, i.e. mobility synchronisation. Each item corresponds to a column in the dataset.

\begin{description}
    \item[year:] Year when the location information was saved in the database.
    \item[day:]  Number between 1 and 365 (or 366 for leap years) where January 1 is day 1. This indicates the day when the out-of-home event was identified.
    \item[hour:] Integer between 0 and 24 which represents the hour when the Out-of-Home event was identified.
    \item[trips:] Total number of times that the users in a given area left home at a given period of time, i.e., the number of Out-of-home trips. The values were normalised using a min-max normalisation to comply with data sharing policies.
    \item[geo\_code:] Unique codes for local authority districts (LAD) and unitary authorities (UA) in the United Kingdom.
\end{description}

\subsection{Trips to Green Areas}

Description of the data used to analyse the visits to green ares. Each item corresponds to a column in the dataset.

\begin{description}
    \item[year:] Year when the location information was saved in the database.
    \item[week:]  Week number according to the ISO-8601 standard where weeks start on Monday. It is a number between 1 and 52 (or 53 for leap years).
    \item[visits:] Total number of users that presented a trip that contains a green area in the period considered. The values were normalised using a min-max normalisation to comply with data sharing policies.
    \item[geo\_code:] Unique codes for local authority districts (LAD) and unitary authorities (UA) in the United Kingdom.
\end{description}

\subsection{Duration of the Trips}

Description of the data used to analyse the duration of the trips. Each item corresponds to a column in the dataset.

\begin{description}
    \item[year:] Year when the location information was saved in the database.
    \item[week:] Week number according to the ISO-8601 standard where weeks start on Monday. It is a number between 1 and 52 (or 53 for leap years).
    \item[hour\_leave:] Integer between 0 and 24 which represents the hour when the user left their home geofencing area.
    \item[hour\_return:] Integer between 0 and 24 which represents the hour when the user first entered in their home geofencing area.
    \item[trips:] Total number of users that presented that trip. The values were normalised using a min-max normalisation to comply with data sharing policies.
    \item[geo\_code:] Unique codes for local authority districts (LAD) and unitary authorities (UA) in the United Kingdom.
\end{description}

\begin{figure}[!ht]
\centering
\includegraphics[width=0.95\textwidth]{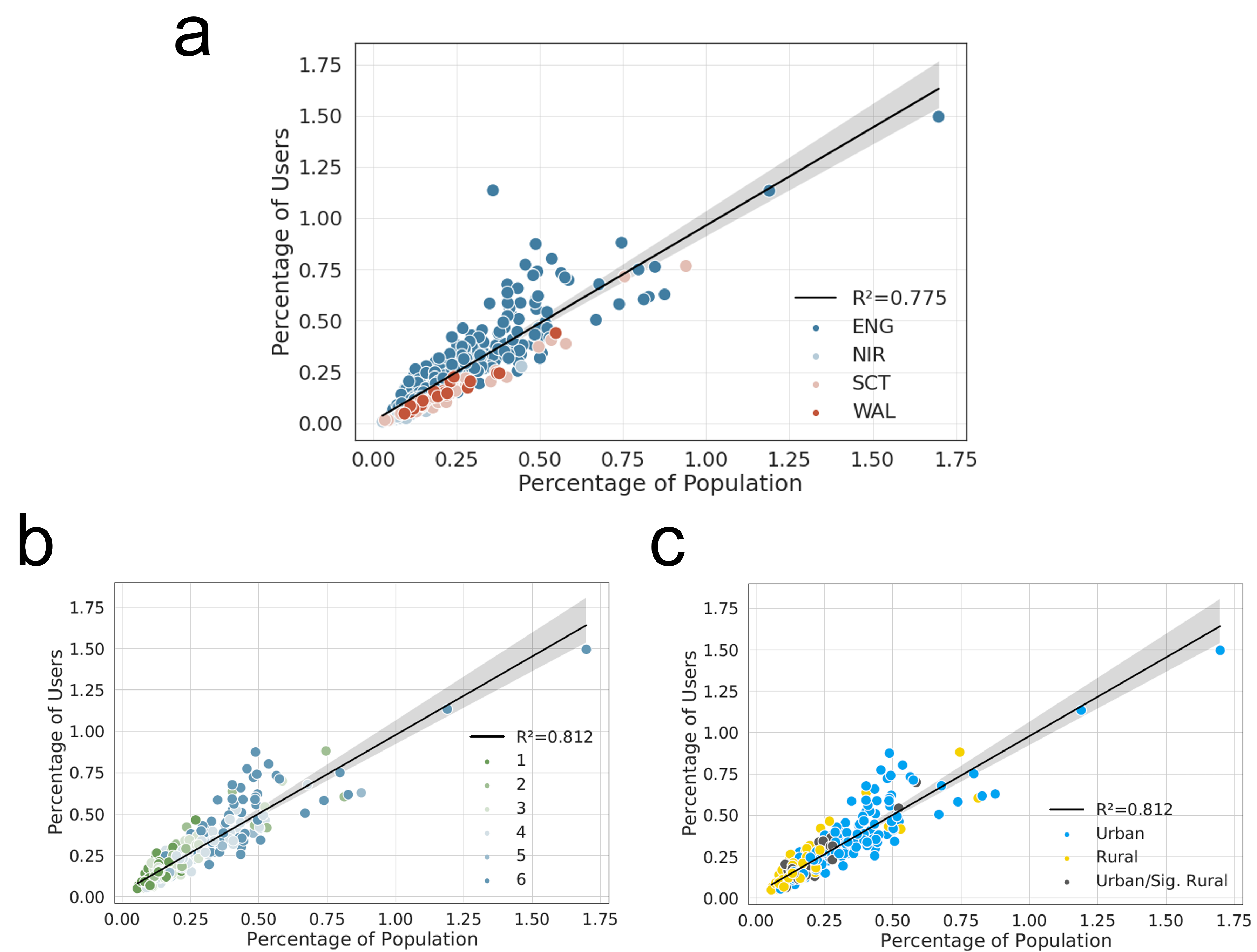}
\caption{Correlation between the percentage of users and the population. (a) Each point in the scatter plot represents a local authority in the UK ($N=404$ local authorities), and the colour indicates its country. For (b) and (c), each point in the scatter plot represents a local authority in England ($N=316$ local authorities), and the colour indicates its level of urbanisation (b) or urban-rural group (c). The x-axis represents the division of the region's population by the sum of the population of all regions. Similarly, the y-axis represents the number of users in a local authority divided by the total number of users in the data. For the results reported, $P < 0.001$ and $CI=95\%$.}
\label{fig:data_validation}
\end{figure}

In \ref{fig:data_validation} each point represents a local authority colour-coded by their country, and the axis represents the number of users and the population as a percentage of the total.  A strong positive correlation between the populations compared is observed in \ref{fig:data_validation} with $r^2$ value of 0.78. Performing the same analysis for English local authorities used in studies related to the level of urbanisation, \ref{fig:data_validation} (b) and (c) shows a stronger correlation of 0.81.

\begin{figure}[!ht]
\centering
\includegraphics[width=0.95\textwidth]{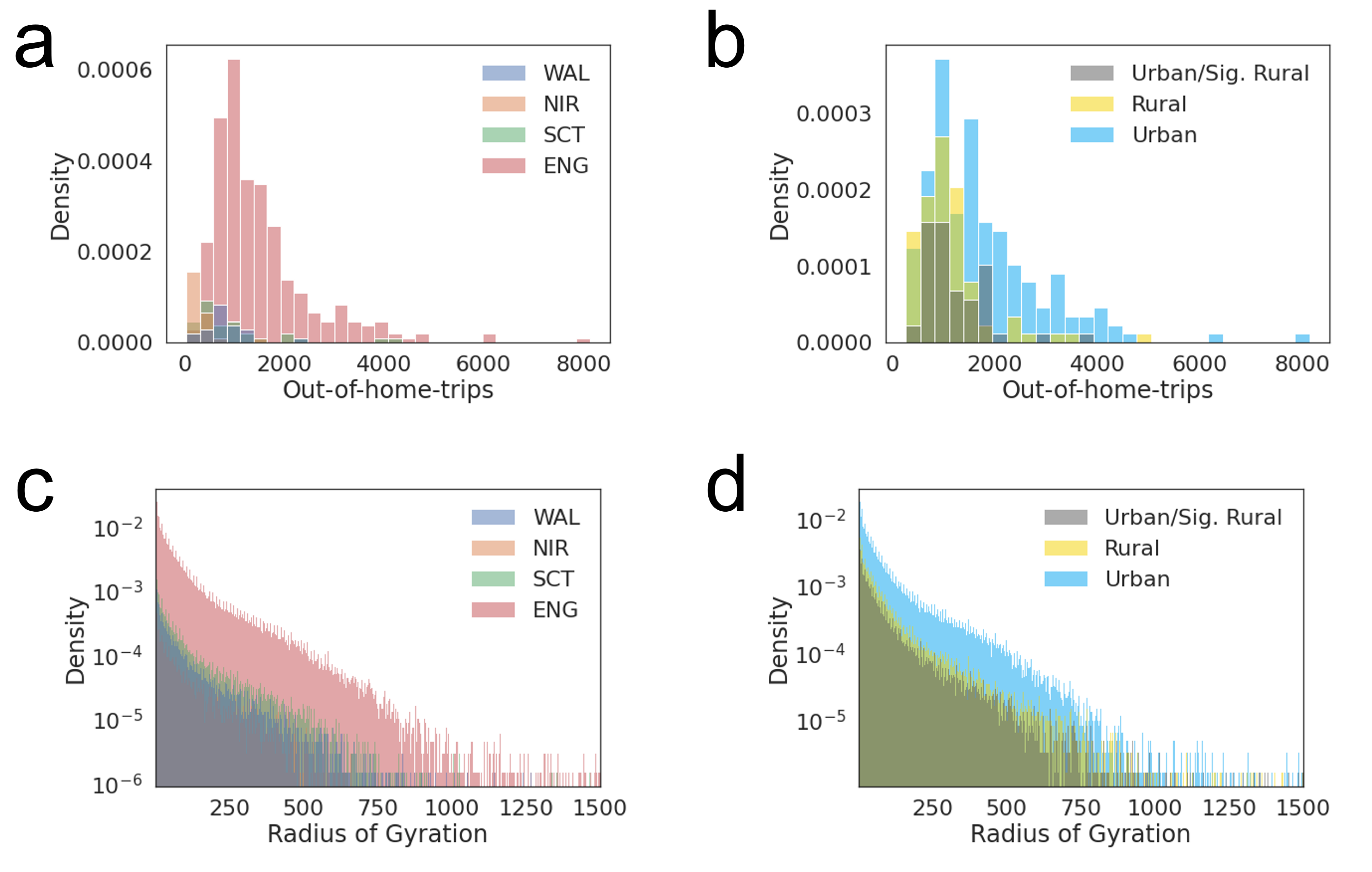}
\caption{Distribution of the number of out-of-home-trips by county (a) and by urban-rural group (b) and the radius of gyration (km) by county (c) and by urban-rural group (d).}
\label{fig:out_hist}
\end{figure}

The dataset contains information regarding the duration of out-of-home trips for users in the UK. The data is aggregated at the local authority level, so no information that allows user identification is included. This data will be used to assess the differences in trip duration before, after and during periods when mobility restrictions were in place.

To further assess the representativeness of the data used, we analysed the data distribution by country and urbanisation level. The results depicted in \ref{fig:out_hist} do not show elements of concerns of a region of the group being able to influence their result over-weighing the distributions. Note that, for the level of urbanisation and socioeconomic studies, we only use data from England due to the lack of a stand data representation for all UK's countries.  Moreover, before the analysis, we filter the data set and remove users with abnormal activity, such as too few logs in the server, lower location accuracy or a large range of motion (e.g. users travelling more than 100km in a day).

\clearpage
\newpage

\section{NS-SEC Classification}
\label{sec:nssec}

The urban-rural classification adopted for England has two levels of classification. The first level classifies areas as urban, rural and urban with significant rural, while the second level classifies the areas into six levels of urbanisation ranging from level 1 to level 6. Level 1 corresponds to rural areas or low urbanisation index, while level 6 describes areas classified as major urban or high-urbanisation indexes. \ref{tab:ns-sec} summarises the NS-SEC classes used in our analysis.

\begin{table}[!ht]
    \centering
	\begin{tabular}{lp{0.6\textwidth}}
        \hline
		{\textbf{Class}} & {\textbf{Description}}\\
        \hline
		NS-SEC 1  & Higher managerial, administrative and professional occupations \\ 
		NS-SEC 2  & Lower managerial, administrative and professional occupations \\
		NS-SEC 3  & Intermediate occupations (e.g. Intermediate clerical, administrative, sales, service, engineering, technical and auxiliary occupations)\\
		NS-SEC 4  & Small employers and own account workers \\
		NS-SEC 5  & Lower supervisory and technical occupations \\
		NS-SEC 6  & Semi-routine occupations \\
		NS-SEC 7  & Routine occupations \\
		NS-SEC 8  & Never worked and long-term unemployed \\
		\hline 
	\end{tabular}
\caption{The National Statistics Socio-economic classification (NS-SEC) classes description.} 
\label{tab:ns-sec}
\end{table}

\section{Spatio-temporal Metrics}
\label{sec:metrics}

\begin{figure}[!ht]
\centering
\includegraphics[width=0.95\textwidth]{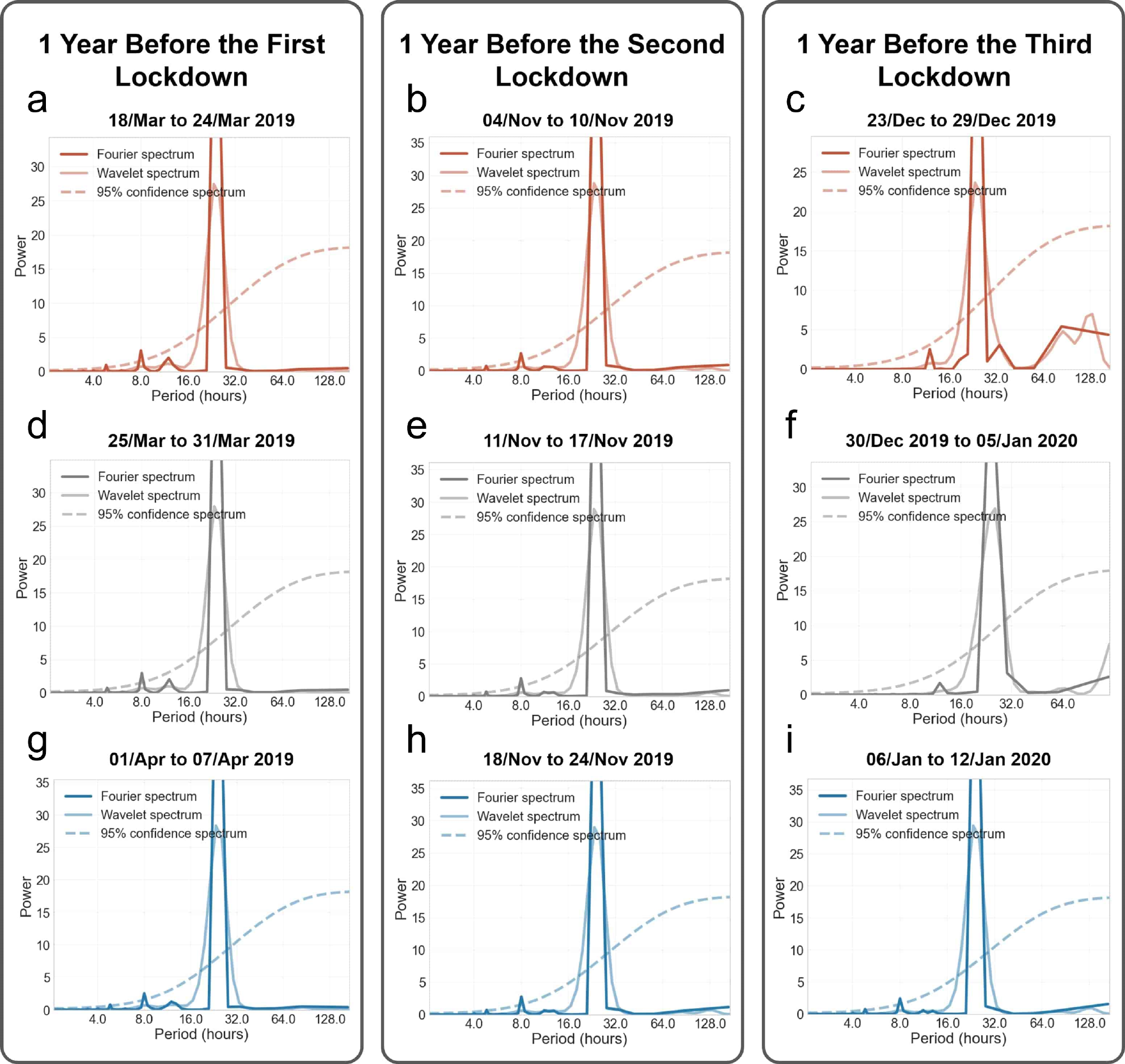}
\caption{Assessment of the mobility synchronisation periods for the baseline year (i.e. 2019 and January 2020).  (a), (b), and (c) represent the patterns for the baseline period corresponding to the week before the announcement of the lockdowns. (d), (e), and (f), depict the baseline week's patterns when the lockdowns were announced. Lastly, (g), (h), and (i) show the baseline patterns for the week corresponding to the second week of lockdown. The synchronisation periods above the 95\% confidence interval is based on the global wavelet and Fourier for the \emph{out-of-home} trips spectra on ea, ch period. Note that, in the baseline, the most significant periods are t, o the,  and 24hrs. However, since the 24hrs components dominate the other, it was not considered for calculating the mobility synchronisation metric.}
\label{fig:sync_fourier_2019}
\end{figure}

\begin{figure}[!ht]
\centering
\includegraphics[width=0.95\textwidth]{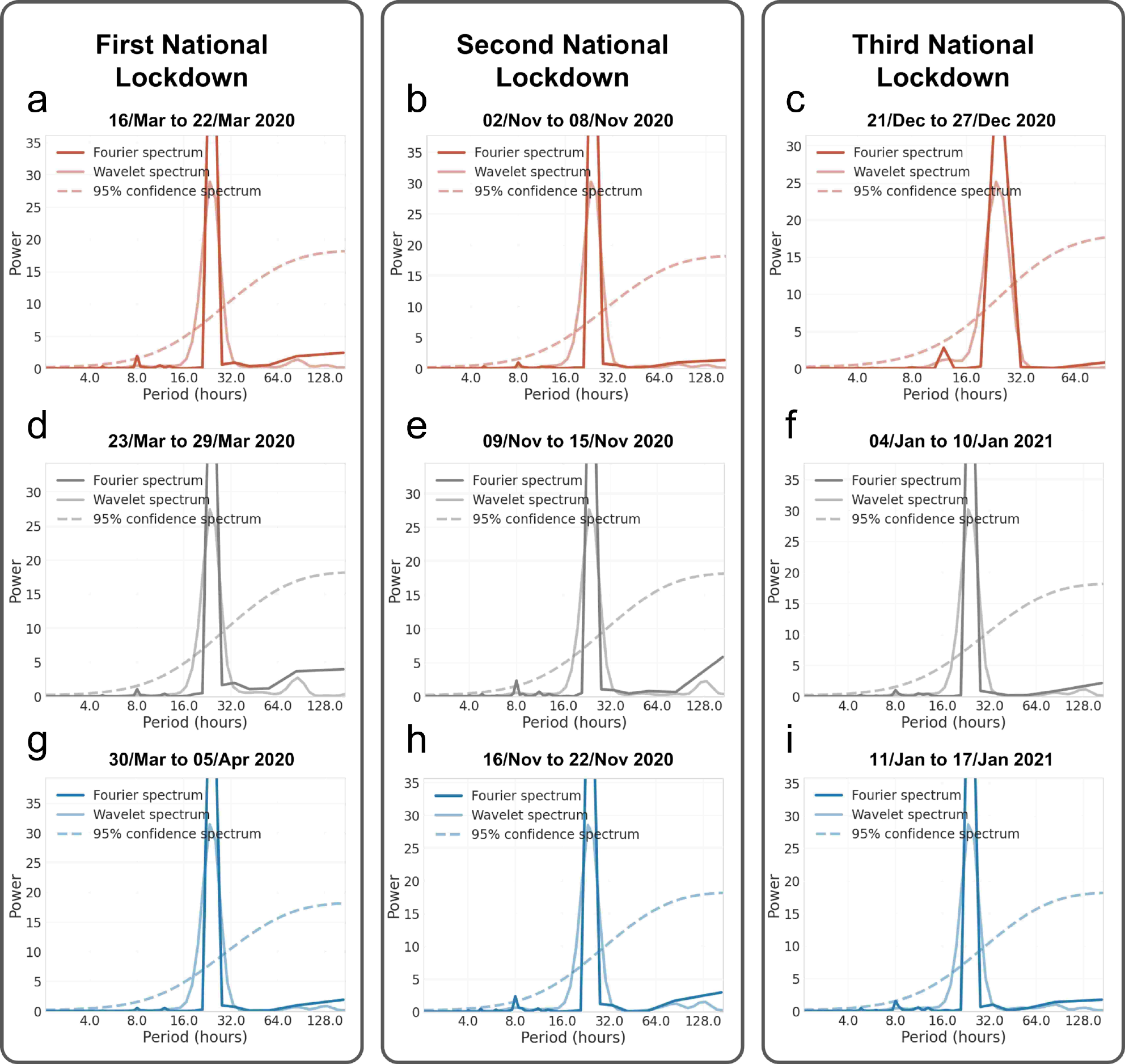}
\caption{Assessment of the mobility synchronisation in the UK before the national lockdowns (a), (b), and (c), in the first week of the lockdowns (d), (e), and (f), and in the second week of the lockdowns (g) and (h)  The synchronisation periods above the 95\% confidence interval is based on the global wavelet and Fourier for the \emph{out-of-home} trips spectra on each period. Note that the synchronisation pattern was mostly changed during the lockdown period as (d) and (e) show a reduction in the spike at 8hrs mark when compared to the period before first lockdown (a).}
\label{fig:sync_fourier_2020}
\end{figure}

\begin{figure}[!ht]
\centering
\includegraphics[width=0.95\textwidth]{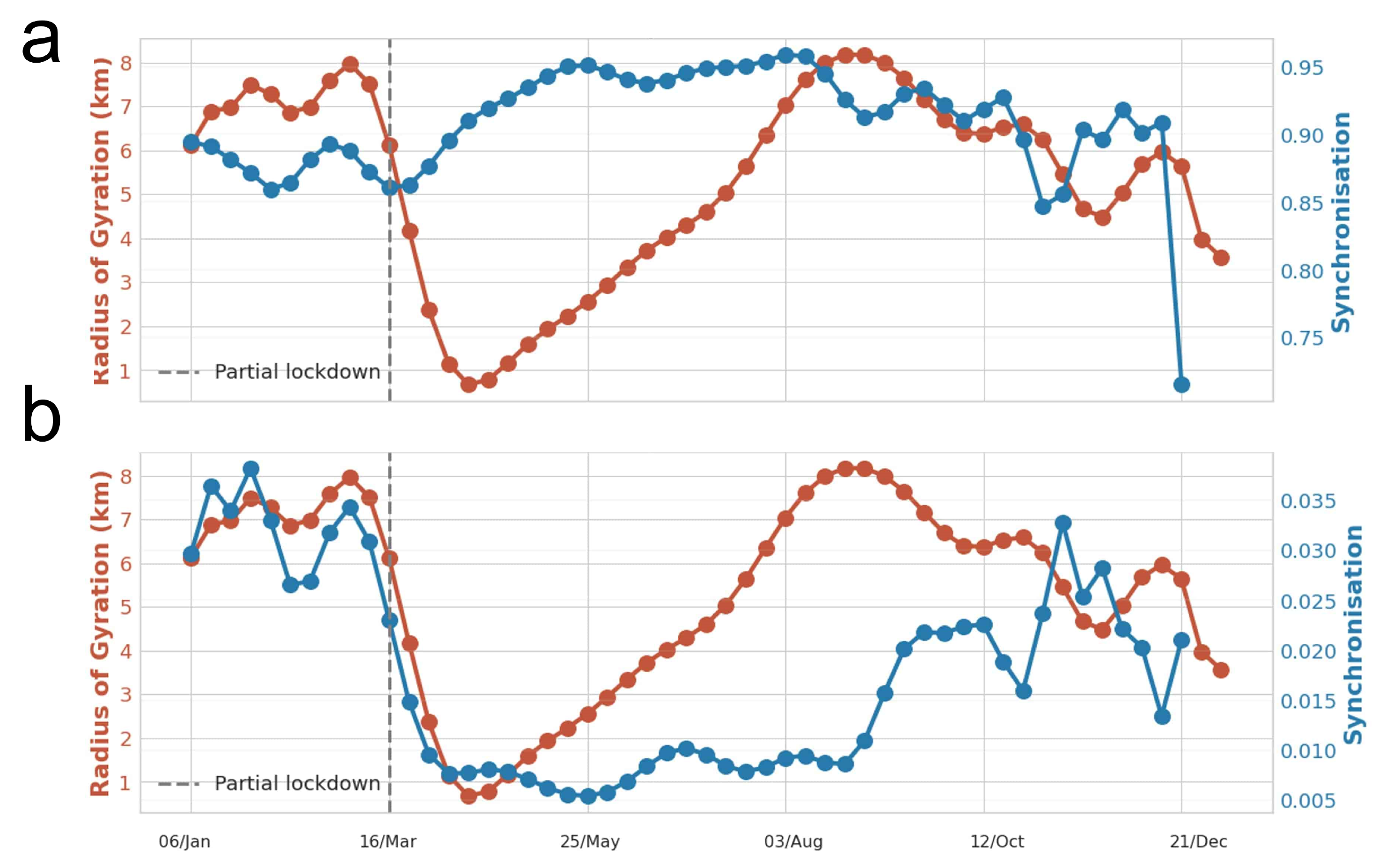}
\caption{Comparison between the Spatio-temporal dimensions of mobility. (a) shows the calculation of the mobility synchronisation metric using the 24hrs components, while (b) depicts the behaviour of this metric without it. Note the similarities in the trends of both metrics in (b)) for the period before the first lockdown.}
\label{fig:sync_24hrs_component}
\end{figure}

\clearpage
\newpage
\section{Mobility Patterns in England}
\label{sec:eng}

\begin{figure}[!ht]
\centering
\includegraphics[width=0.9\textwidth]{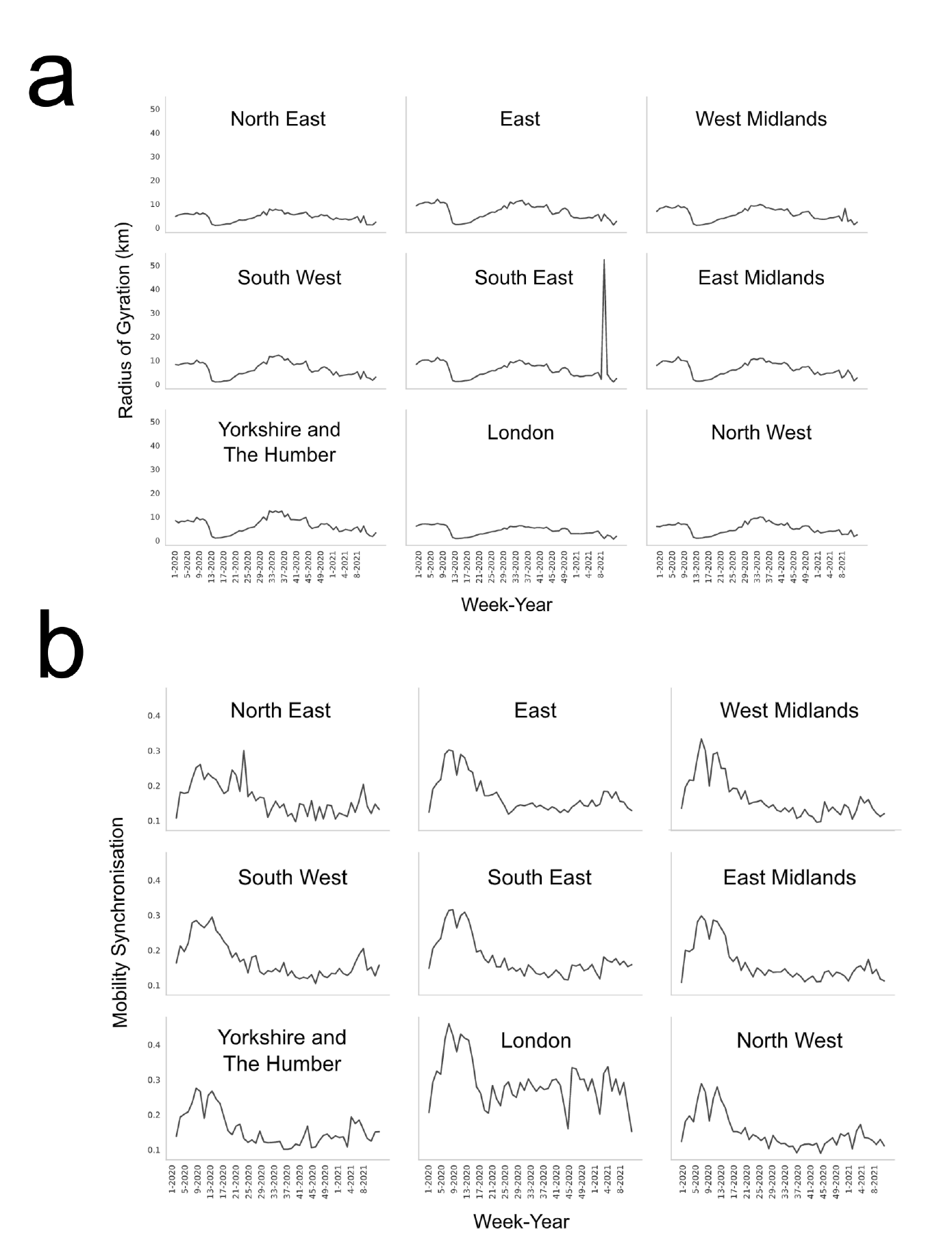}
\caption{Radius of gyration and mobility synchronisation time series for the regions in England. Period from week 1 2020 until week 8 2021.}
\label{fig:trends_eng_radius}
\end{figure}

\clearpage
\newpage
\section{Mobility Patterns in Northern Ireland}
\label{sec:nir}

\begin{figure}[!ht]
\centering
\includegraphics[width=0.9\textwidth]{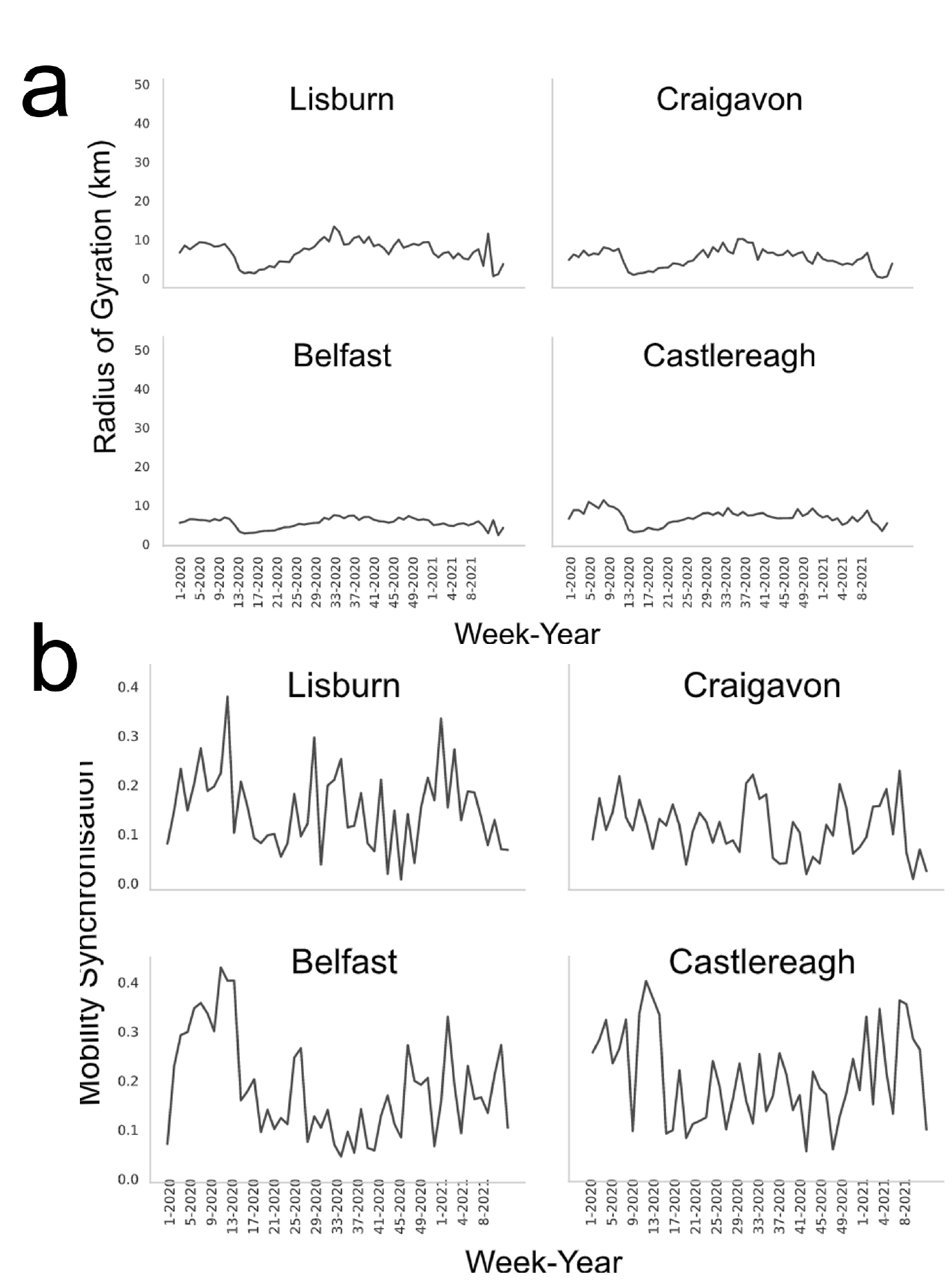}
\caption{Radius of gyration and mobility synchronisation time series for the studied areas in Northern Ireland. Period from week 1 2020 until week 8 2021.}
\label{fig:trends_nir_radius}
\end{figure}

\clearpage
\newpage
\section{Mobility Patterns in Scotland}
\label{sec:sct}

\begin{figure}[!ht]
\centering
\includegraphics[width=0.95\textwidth]{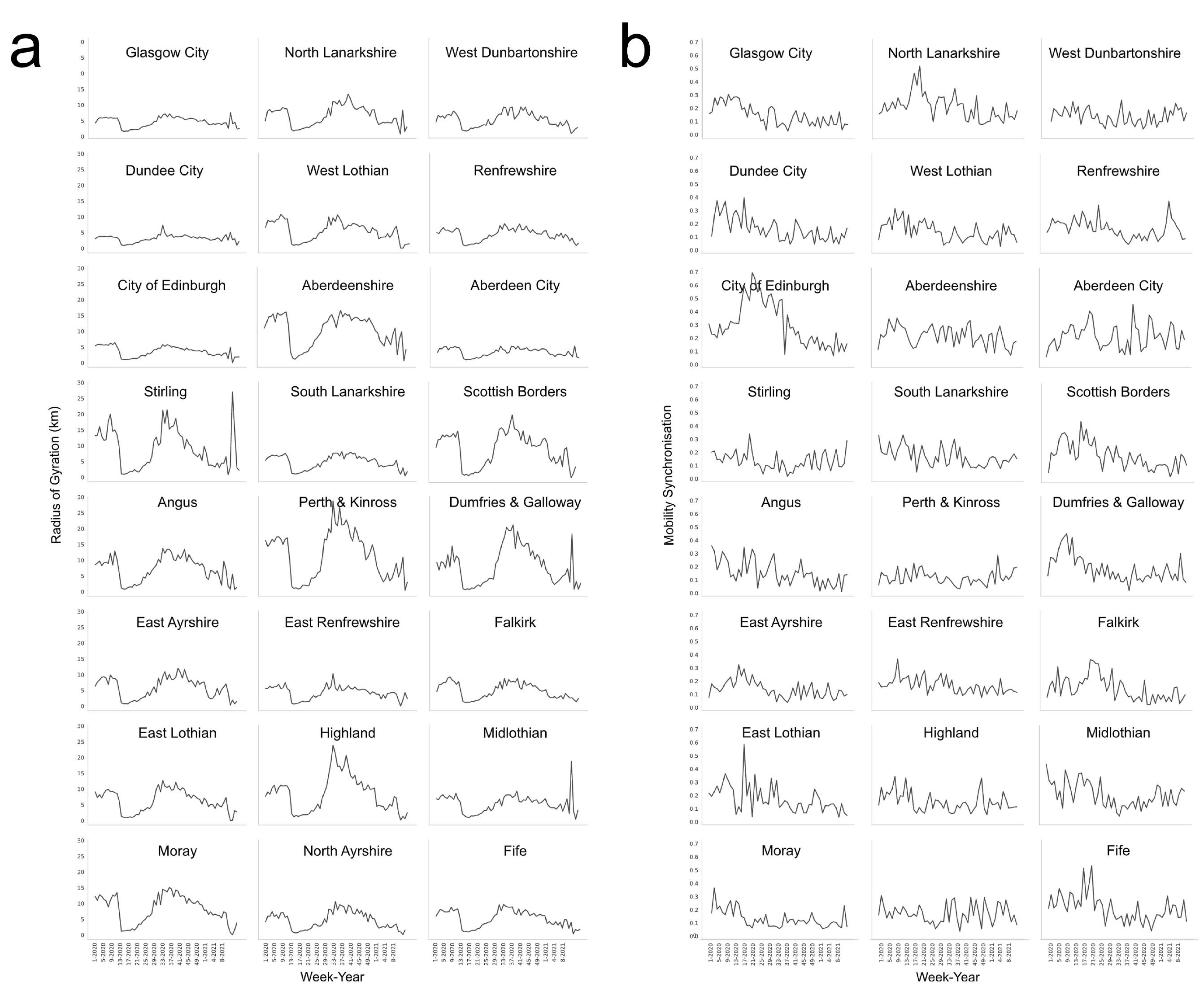}
\caption{Radius of gyration and mobility synchronisation time series for the studied areas in Scotland. Period from week 1 2020 until week 8 2021.}
\label{fig:trends_sct_radius}
\end{figure}

\clearpage
\newpage
\section{Mobility Patterns in Wales}
\label{sec:wal}

\begin{figure}[!ht]
\centering
\includegraphics[width=0.95\textwidth]{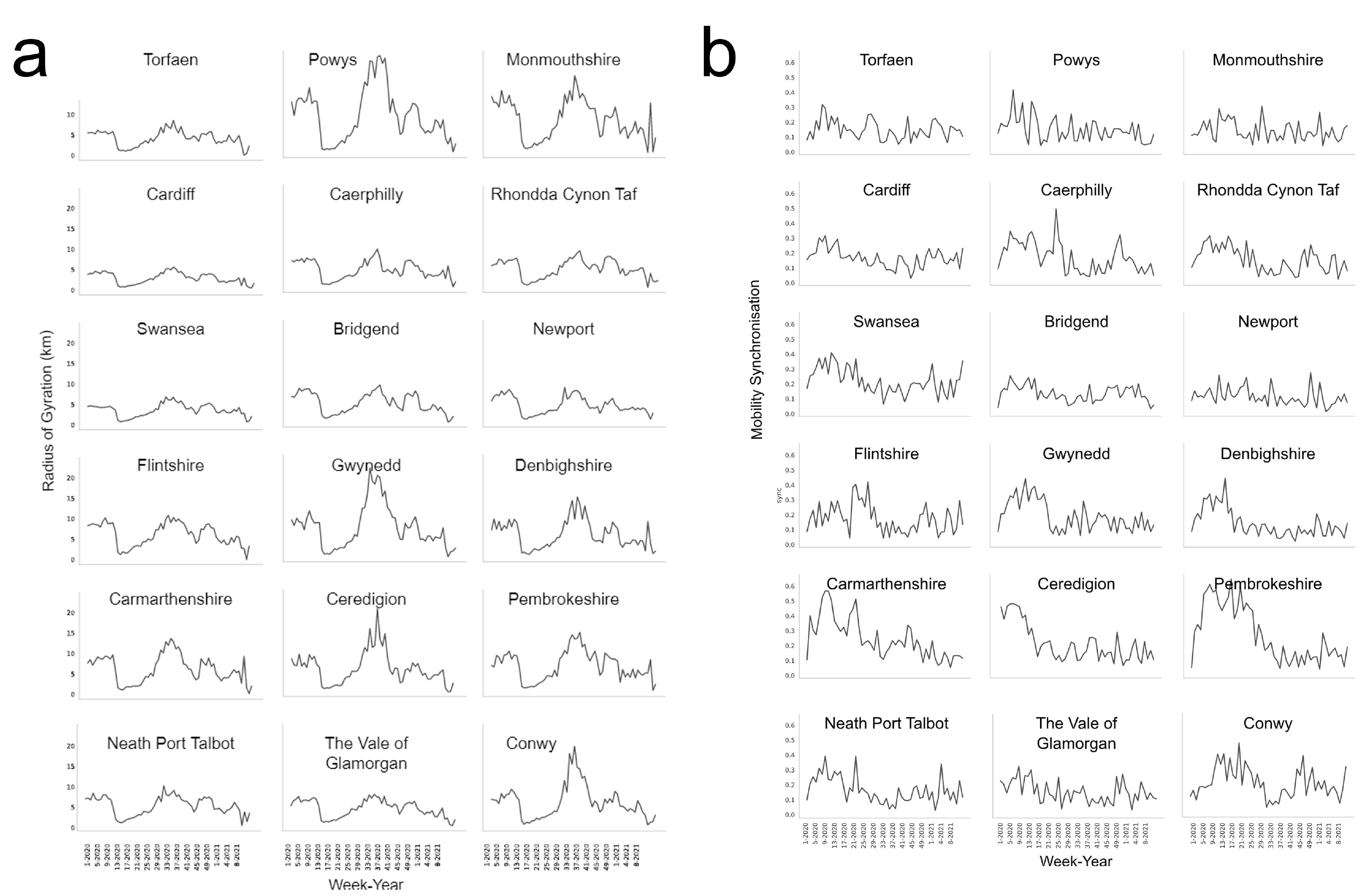}
\caption{Radius of gyration and mobility synchronisation time series for the studied areas in Wales. Period from week 1 2020 until week 8 2021.}
\label{fig:trends_wal_radius}
\end{figure}

\end{document}